\documentclass[twocolumn,prX]{revtex4} 
\usepackage{epsfig}
\begin{document}
\newcommand{\pint}{\Pi\hspace{-4mm}\int}
\renewcommand{\Im}{{\rm Im}}

\title{Landau-Zener Probability Reviewed}
\author{C. Valencia*$^{\dagger}$, D.E. Jaramillo*\\{\em *Instituto de F\'\i sica, Universidad de Antioquia, A.A. 12 
26,Medell\'\i n, Colombia}
\\{\em $^\dagger$Instituto Tecnol\'ogico Metropolitano, Calle 73 No 76A -354,  Medell\'\i n - Colombia }}

\begin{abstract}
We examine the  survival probability for neutrino propagation through matter with variable density. We present a new 
method to calculate the level-crossing probability that differs from Landau's method by constant factor, which is  
relevant in the interpretation of neutrino flux from supernova explosion.\\

\end{abstract}

\maketitle
\section{Introduction}
The study of neutrino masses and mixing is one of the most interesting issues in particle physics which has also 
considerable impact on astrophysical and 
cosmological problem.  Looking for evidence of mixing neutrino flavors 
during its propagation is one method to detect massive neutrinos. If neutrinos propagate 
through matter, mixing effects can be enhanced. The electrons in the background matter 
induce the mass to the electron neutrino trough a charged current. 
In non-uniform medium density changes on the way of neutrinos therefore the mixing angle changes during propagation and   
the eigenstates of the Hamiltonian are no more eigenstates of propagation. Transitions between mass eigenstates can  
occur. The level crossing  probability is known as the Landau-Zener probability \cite{1}.   If density changes slowly 
enough those transition can be neglected so the  
mass eigenstates propagates independently, as it does in the vacuum or in a uniform medium. This is called the adiabatic 
condition. The solar neutrino conversion is correctly described with the adiabatic condition with accuracy of $10^{-7}$ 
\cite{smir}. If the density changes rapidly like inside supernovas the adiabatic condition is not satisfied, then the 
probability of transition between the mass eigenstates becomes relevant. 

In this paper we focus our attention in the deduction of the level crossing probability  expanding the temporal evolution 
operator, we found an general expression for 
this probability and we arrived to the usual one  taking  the first term in the perturbation expansion. In section \ref{2} 
we briefly review the basic 
elements for describing neutrino oscillations in a medium, the standard classic probability is derived from a geometrical 
picture. In section \ref{3} We develop a 
perturbation method to find the temporal evolution which allow us  to find the level crossing probability. We found  that 
it differs from  Landau-Zener 
probability  by a factor $\pi^2/4$.


\section{Formalism }\label{2}
In the standard model of neutrinos \cite{smir2}  with $\theta_{13}\sim 0$ a neutrino state propagating in the matter is 
assumed to be a linear combination of 
the flavor states $|\nu_e\rangle$ and $|\nu_\alpha\rangle$ 

\[ |\nu(t)\rangle =\nu_e(t)|\nu_e\rangle +\nu_\alpha(t)|\nu_\alpha\rangle \] 
with $|\nu_\alpha\rangle$ being a determined linear combination of $|\nu_\mu\rangle$ and $|\nu_\tau\rangle$. 

The two-neutrino system propagating in matter obeys the   Schrodinger equation
\begin{equation}\label{1}
i\frac{d}{dt}\nu=
H\nu,
\end{equation}
with $\nu=(\nu_e,\nu_\alpha)^T$. Using  the Pauli spin matrices  $\sigma_i$ in the ultra relativistic approximation the 
Hamiltonian can be written as \cite{kim}
\begin{equation}\label{3}
H=(\overline m+\Delta_0a)\sigma_0-\Delta_0(e^{-2i\sigma_2\theta}+ a)\sigma_3,
\end{equation}
where $2\Delta_0 a=\sqrt{2}\,n_eG_F$,  $\overline m=(m^2_2+m^2_1)/4E$ and $\Delta_0=(m_2^2-m_1^2)/4E>0$. 

In  the matter basis, $\nu_m= e^{-i\phi\sigma_2}\nu$, the  Hamiltonian is diagonalized to
\begin{equation}\label{6}
e^{-i\phi\sigma_2}He^{i\phi\sigma_2}=(\overline m+\Delta_0a)\sigma_0-\Delta\sigma_3
\end{equation}
where 
\begin{equation}\label{delta}
\Delta=\Delta_0\sqrt{1-2\cos2\theta a+a^2},
\end{equation}
and
\begin{equation}\label{phi}
\displaystyle{\cot2\phi=\frac{\cos2\theta-a}{\sin2\theta}}
\end{equation}
which give us two eigenvalues 
\begin{equation}
E_\pm=\overline m+\Delta_0\left(a\pm\sqrt{1-2\cos2\theta a+a^2}\right)
\end{equation}
associated with the effective masses $m_\pm= \sqrt{4EE_\pm}$. 


\section{Semi-classic Probability}\label{3}

Plotting $E_\pm$ with respect to $a$ we find two hyperbolas with the asymptotic behavior
\begin{equation}
 \omega_\pm=\overline m+\Delta_0\Big(a\pm (a -\cos2\theta)\Big).
\end{equation}
The differences between the curves and the asymptotes 
satisfy
\begin{equation} 
\frac{|E_+-\omega_+|}{|E_--\omega_+|}=\frac{1-\cos2\phi}{1+\cos2\phi}=\frac{\sin^2\phi}{\cos^2\phi}
\end{equation}  
so the asymptotes $\omega_\pm=\cos^2\phi E_\pm+\sin^2\phi E_\mp$ represent the mean value of the squared mass in the 
flavor states. 
\begin{figure}[ht]
\begin{center}
\epsfig{file=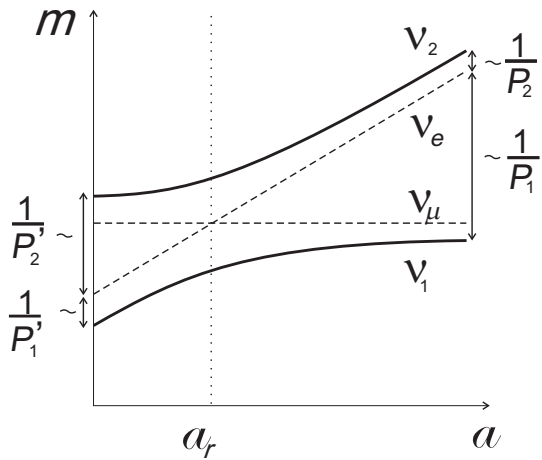, height=6cm ,angle=0}
\caption{Evolution of the probability}\label{masses}
\end{center}
\end{figure}
Furthermore the probability of finding the eigenstate in some flavor is given by how close the hyperboles  are from the 
asymptotes. We can interpret  FIG. \ref{masses} in the classical  way. Let us suppose that $N$ electronic neutrinos  
 are produced inside matter, classically there is $N_1=N P_{1}$  neutrinos of mass $m_+$ and
$N_2=N P_{2}$  neutrinos of mass $m_-$. When they go  into the vacuum they are going detected like  $N_e 
=P^0_{1}N_1+P^0_{2}N_2$ 
neutrinos of electronic type. Then the survival probability is
\begin{equation}
P_{\nu_e\rightarrow\nu_e}=\frac{N_e}{N}= P^0_{1}P_{1}+P^0_{2}P_{2}=\frac{1}{2}(1+\cos2\theta\cos2\phi).
\end{equation}  
Actually the neutrinos of mass $m_+$ travelling trough matter can be converted into neutrinos of mass $m_-$  and 
vice-versa because a quantum tunneling effect. The number of conversions must be proportional to the difference $N_1-N_2$, 
so when they travel in the vacuum there 
will be    $N^0_1=N_1+P(N_2-N_1)$ neutrinos of mass $m_+$ and $N^0_2=N_2+P(N_1-N_2)$ neutrinos of mass  $m_-$, where $P$ 
is the conversion probability. The number of detected electronic neutrinos is  $N_e =P^0_{1}N^0_1+P^0_{2}N^0_2$ and the
probability for detecting a neutrino electronic is now
\begin{eqnarray}\label{pclass}
P_{\nu_e\rightarrow\nu_e}&=& P^0_{1}P_{1}+P^0_{2}P_{2}-P(P_1-P_2)(P^0_1-P^0_2)\nonumber\\&=&\frac{1}{2}\Big(1+(1-2P)
\cos2\theta\cos2\phi\Big).
\end{eqnarray}  
The conversion probability $P$ is known the Landau-Zener probability 
\section{Quantum probability}

Now let us calculate the Landau-Zener probability from the Schrodinger equation for the neutrino system.
Terms proportional to  $\sigma _0$ in (\ref{6}) contribute only with an overall phase physically meaningless, so we can 
drop it. 
When neutrinos are produced inside matter the mixing angle changes if the  density  is a function of the 
position. The angle $\phi$ depends on time while the neutrino traveling in matter. The Schrodinger equation in  the matter 
eigenstates now read 
\begin{equation}\label{h1}
ie^{-i\phi\sigma_2}\frac{d}{dt}e^{i\phi\sigma_2}\nu_m=-\Delta\sigma_3\nu_m
\end{equation}
that is
\begin{equation}\label{h2}
i\frac{d}{dt}\nu_m=(\dot\phi\sigma_2-\Delta\sigma_3)\nu_m\equiv H_e \nu_m .
\end{equation}
From (\ref{h2}) $|\dot \phi|$ determine the energy transition between the two eigenstates and $\Delta$ give the gap 
between levels. If
\begin{equation}
\frac{|\dot\phi|}{\Delta}\ll 1,
\end{equation} 
the off-diagonal terms of the effective Hamiltonian $H_e$ can be neglected and the system of equations for the eigenstates 
decouple. This is the condition of adiabaticity.
For non-adiabatic limit we can not decouple  the   neutrino system. 

The survival probability for electronic neutrino is 
\begin{equation}\label{p1}
P_{\nu_e\rightarrow\nu_e}(t)=\left|\chi_\theta ^\dagger U(t)\chi_\phi \right|^2
\end{equation} 
where $\chi_\alpha= (\cos\alpha,\sin\alpha)^T$ are the components of the neutrino electronic in the basis of the 
Hamiltonian eigenstates.  $U(t)$ is  the temporal evolution operator which satisfy the Schrodinger equation

\begin{equation}\label{h3}
i\frac{d}{dt}U(t)=H_e(t)U(t).
\end{equation}
Because of unitary $U(t)$ can be written as
 
\begin{equation}\label{us}U(t)=a_0(t)\sigma_0+i \vec a (t)\cdot \vec\sigma ,
\end{equation}
where the $a_i$ are real and $\sum_i a_i=1$. 
The probability (\ref{p1}) in function of this parameters is written as\cite{Park} 
\begin{widetext}
\begin{eqnarray}\nonumber P_{\nu_e\rightarrow\nu_e}(t)
=\frac{1}{2} \Big( 1+\cos{2\theta}\cos{2\phi}(a_0^2-a_1^2-a_2^2+a_3^2)+\cos{2\theta}\sin{2\phi}(a_1a_3-a_0a_2)\\
+\sin{2\theta}\sin{2\phi}(a_0^2+a_1^2-a_2^2-a_3^2)+\sin{2\theta}\cos{2\phi}(a_1a_3+a_0a_2)\Big) . \label{aaaa} 
\end{eqnarray}
\end{widetext}
The $a_i$ coefficients can be found solving equation (\ref{h3}) which can be written in a differential form
\begin{equation}\label{h4}
U(t+dt)=\Big(1-iH_e(t)dt\Big)U(t),
\end{equation}
with the condition  $U(0)=1$. From (\ref{h4}) it is straightforward to find that
\begin{eqnarray}\label{u}
U(t)=\lim_{N\rightarrow \infty}\prod_{k=0}^N \Big(1-iH_e(kt/N)dt\Big)
\equiv \Pi\hspace{-4mm}\int_0^t e^{-iH_e(t)dt}. 
\end{eqnarray}
If $[H(t_1),H(t_2)]=0$ for any pair $(t_1, t_2)$   trivially
\begin{equation}\pint_0^t e^{-iH dt}= e^{-i\int_0^t H(t)dt}.\end{equation}
If the Hamiltonian does not commute for different times we can do perturbation theory splitting the effective Hamiltonian 
in a no perturbed and perturbation parts,  $H_{e}=H_0+H_1$, 
 in our case
\begin{equation}\label{hh}
H_0=-\Delta\sigma_3,\;\; H_1=\dot\phi\sigma_2.
\end{equation} 
Assuming $\langle H_0\rangle \gg \langle H_1\rangle$ we  expand  
 (\ref{u}) as
\begin{widetext}
\begin{eqnarray*}
U(t)&=&.. \Big(e^{-iH_0(t-dt)dt}-iH_1(t-dt)dt\Big)\Big(e^{-iH_0(t)dt}-iH_1(t)dt\Big)\Big(e^{-iH_0(t+dt)dt}-iH_1(t+dt)dt
\Big) .. \\
&=& e^{-i\int_0^tH_0dt} \\[2mm]
&&-i\int_0^tdt_1e^{-i\int_0^{t_1}H_0dt}H_1(t_1)e^{-i\int_{t_1}^{t}H_0dt}\\
&&-\int_0^tdt_2\int_0^{t_2} dt_1e^{-i\int_0^{t_1}H_0dt}H_1(t_1)e^{-i\int_{t_1}^{t_2}H_0dt}H_1(t_2)e^{-i\int_
{t_2}^{t}H_0dt} .
\end{eqnarray*}
\end{widetext}
This equation can  be graphically represented  as 


\begin{picture}(100,50)
\put(0,20){$U(t)=$}
\put(32,23){\line(1,0){25}}
\put(65,20){+}
\put(80,21){\line(4,1){25}}
\put(102,25){$\times$}
\put(105,8){$^1$}
\put(170,25){$\times$}
\put(171,8){$^1$}
\put(192,20){$\times$}
\put(196,6){$^2$}
\put(106,27.5){\line(4,-1){25}}
\put(140,20){+}
\put(153,22){\line(4,1){20}}
\put(175,27){\line(4,-1){20}}
\put(195,22){\line(4,1){20}}
\put(220,20){+ ..}
\end{picture}

with the Feynman rules

\begin{picture}(300,50)
\put(5,13){\line(1,0){40}}
\put(55,10){$=e^{-i\int H_0dt},$}
\put(122,11){$\times\;= \displaystyle{\int dt_j .... \,(-iH_1(t_j))}.$}
\put(124,0){$^j$}
\end{picture}


Using (\ref{hh}) the time evolution operator can be expressed  as

\begin{eqnarray}\nonumber
U(t)= \left(\sum_{n=0}^{\infty} (-i\sigma_2)^n \int \prod_{j=1}^n dt_j\dot\phi(t_j) e^{2i\sigma_3\int_{0}^{t_j}\Delta 
dt(-)^{n+j}} \right)\\
\times e^{-i\sigma_3\int_0^t\Delta dt}\;\;\;\label{U}
\end{eqnarray}
which can be parametrized as 

\begin{equation}
U(t)=(\cos\lambda +i\sigma_2 \sin\lambda e^{i\beta\sigma_3})e^{i\alpha\sigma_3} ,
\end{equation}
where $\alpha$ is a monotonous function of time and $\beta$ depends on the time to reach the vacuum. Comparing with (\ref
{us})   we find for  the fully  averaged probability  (\ref{aaaa}), over the time of  production and detection, is
\begin{eqnarray}\langle P_{\nu_e\rightarrow\nu_e}\rangle
=\frac{1}{2} \Big( 1+\cos{2\theta}\cos{2\phi}(1-2\sin^2\lambda)
\Big) .
\end{eqnarray}
Comparing with  (\ref{pclass}) we can see that the probability conversion is given by $P=\sin^2\lambda$, which is the 
modulo squared of the $\sigma_2$ coefficient in (\ref{U}), that is

\begin{equation}\label{p0}
P=\left|\sum_{n=0}^{\infty}(-1)^n \int \prod_{j=1}^{2n+1} dt_j\dot\phi(t_j) e^{-2i\int_{0}^{t_j}\Delta dt(-)^{j}} \right|^2.
\end{equation}
This is an exact expression for the Landau-Zener probability.
At lowest order in $\dot\phi$, the Landau-Zener probability is 
\begin{equation}\label{p1}
P_{LZ}=\left|\int_0^t dt_1\dot\phi(t_1) e^{2i\int_0^{t_1}\Delta dt} \right|^2. 
\end{equation}
From (\ref{phi}) we obtain
\begin{equation}
\dot\phi= \frac{\dot a\sin2\theta}{2(1-2a\cos2\theta+a^2)} .
\end{equation}
 Considering that the main contribution is near the resonance region,  $a=\cos2\theta$ and assuming the neutrinos are 
produced above this region we can extend the limits of the integral in (\ref{p1}) over all $\tilde a= a-\cos2\theta$,
\begin{equation}\label{h8}
\int_0^t dt_1\dot\phi(t_1) e^{2i\int_0^{t_1}\Delta dt}\simeq\frac{1}{2}\int_{-\infty}^\infty\frac{\sin2\theta e^{iI(\tilde 
a)}}{\tilde a^2+\sin^22\theta}\,d\tilde a
\end{equation}
where 
\begin{equation}\label{I}
I(\tilde a)=2\Delta_0\int^{\tilde a}\sqrt{\tilde a^2+\sin^22\theta}\frac{d\tilde a}{\dot a}.\end{equation}
The integral in (\ref{h8}) has poles in $\tilde a=\pm i\sin2\theta$. This integral is calculated to give
\begin{equation}
P_{LZ}=\frac{\pi^2}{4}e^{-2|\Im I(i\sin2\theta)|} .
\end{equation}
To find (\ref{I}) we need to know the functional form of $a$.
For example  assuming $\dot a $ constant we have
\begin{equation}
 I(i\sin2\theta )=\frac{2\Delta_0}{\dot a}\frac{\sin^22\theta}{2}\ln(i\sin 2\theta),
\end{equation}
and 
 \begin{equation}\label{plz}
P_{LZ}=\frac{\pi^2}{4}e^{-\gamma\pi/2}
\end{equation}
where
 \begin{equation}
\gamma=\left.\frac{\Delta}{\dot \phi}\right|_{a=\cos\theta}
\end{equation}
is the adiabatic parameter.  
Probabilities for other density distribution can be found in the literature \cite{kuo}. 

In the usual expression for Landau-Zener probability  $P_{LZ} \to 1$ when $\gamma\to 0$. It seems that  (\ref{plz}) is not 
correct because at this limit  $P_{LZ} \to \pi^2/4$ for us.  But in this situation the perturbation approach (\ref{p1}) is 
not valid and we need to take the  expression (\ref{p0}).

\section{Conclusions}

In this paper we have reviewed the  Landau-Zener probability starting from standard approach and introducing a 
perturbation method to solve the temporal evolution operator. We found that our expression differs from the standard one 
by a multiplicative factor $\pi^2/4\sim2.6$ which at the present experimental resolution is irrelevant, but in the 
interpretation of the neutrino flux from  supernova explosion \cite{zulu} could be very important correction because of 
the non adiabatic neutrino propagation. 


\end{document}